\documentclass[aip,jcp,reprint,floatfix]{revtex4-1}
\usepackage{amsfonts,amssymb,amsmath,array}
\usepackage{dcolumn}
\usepackage{bm}
\usepackage[final]{graphicx}
\usepackage{graphicx}
\usepackage{epstopdf}

\usepackage[usenames]{color} 

\usepackage{appendix}

\begin{document}

\title{A microfluidic method for passive trapping of sperms in microstructures}

\author{Binita Nath}
\affiliation{ISC-CNR, Institute for Complex Systems, Piazzale A. Moro 2, I-00185 Rome, Italy}
\affiliation{Dipartimento di Fisica, Sapienza Universit\`a di Roma, Piazzale A. Moro 2, I-00185, Rome, Italy}

\author{Lorenzo Caprini}
\email{lorenzo.caprini@gssi.it, lorenzo.caprini@hhu.de}
\affiliation{Heinrich-Heine-Universität Düsseldorf, Institut für Theoretische Physik II - Soft Matter, 
D-40225 Düsseldorf, Germany}

\author{Claudio Maggi}
\affiliation{NANOTEC-CNR, Institute of Nanotechnology, Soft and Living Matter Laboratory, c/o Dipt. di Fisica, Sapienza Università di Roma, Piazzale A. Moro 2, I-00185, Rome, Italy.}

\author{Alessandra Zizzari}
\affiliation{~NANOTEC-CNR, Institute of Nanotechnology, c/o Campus Ecotekne, University of Salento, Via Monteroni, I-73100, Lecce, Italy.}

\author{Valentina Arima}
\affiliation{~NANOTEC-CNR, Institute of Nanotechnology, c/o Campus Ecotekne, University of Salento, Via Monteroni, I-73100, Lecce, Italy.}

\author{Ilenia Viola}
\affiliation{NANOTEC-CNR, Institute of Nanotechnology, Soft and Living Matter Laboratory, c/o Dipt. di Fisica, Sapienza Università di Roma, Piazzale A. Moro 2, I-00185, Rome, Italy.}

\author{Roberto Di Leonardo}
\affiliation{Dipartimento di Fisica, Sapienza Universit\`a di Roma, Piazzale A. Moro 2, I-00185, Rome, Italy}
\affiliation{NANOTEC-CNR, Institute of Nanotechnology, Soft and Living Matter Laboratory, c/o Dipt. di Fisica, Sapienza Università di Roma, Piazzale A. Moro 2, I-00185, Rome, Italy.}

\author{Andrea Puglisi}
\affiliation{ISC-CNR, Institute for Complex Systems, Piazzale A. Moro 2, I-00185 Rome, Italy}
\affiliation{Dipartimento di Fisica, Sapienza Universit\`a di Roma, Piazzale A. Moro 2, I-00185, Rome, Italy}

\newcommand*{\csperm}{\,\rotatebox[origin=c]{90}{{\large $\sqcap$}}\,}
\newcommand*{\rsperm}{\,\rotatebox[origin=c]{90}{{\large $\cap$}}\,}

\begin{abstract}
Sperm motility is a prerequisite for male
  fertility.  Enhancing the concentration of motile sperms in assisted reproductive
  technologies - for human and animal reproduction - is typically achieved through aggressive methods such as
  centrifugation. Here we propose a passive technique for the amplification of motile sperm concentration, with no
  externally imposed forces or flows. The technique is based upon the disparity between
  probability rates, for motile cells, of entering in and escaping
  from complex structures. The effectiveness of the technique is
  demonstrated in microfluidic experiments with microstructured devices, comparing
  the trapping power in different geometries. In these micro-traps we observe an enhancement of cells’ concentration close to $10$, with a contrast between motile and non-motile increased by a similar factor. Simulations of suitable interacting model sperms in realistic geometries reproduce quantitatively the
  experimental results, extend the range of observations and highlight the ingredients that are key to optimal trap design.
\end{abstract}

\maketitle

\section{Introduction}

It is estimated that the male factor is the origin of roughly half of
the case of fertility problems~\cite{barratt2009human,frey2010male}. Improving the selection
of motile sperms, particularly when they are rare, would be beneficial
for assisted reproduction technologies, for both zootechnics and human
reproduction. These aspects  emphasize  the importance of
scientific research on sperm motility, particularly for mammals, a
field which is certainly blessed by the recent advancements in imaging
techniques, molecular biology and computational analysis~\cite{friedrich2007chemotaxis,friedrich2010high,gaffney2011mammalian,guasto2020flagellar}.

The most successful in-vitro fertilisation techniques
(e.g. Fertilisation in Vitro and Embryo Transfer (FIVET) and
Intracytoplasmic Sperm Injection (ICSI)) are highly aggressive,
particularly for the female partner, and very expensive so that a part
of the population cannot access them~\cite{schill2010andrologia}. Less expensive
techniques, such as Intrauterine Insemination (IUI), is even more
dependent upon the selection of highly motile sperms, a problem which
is typically solved by centrifuge, density gradients and swim-up
techniques~\cite{boomsma2004semen} which can compromise the integrity of cells, mechanically
or by exposition to DNA-disrupting chemical species~\cite{zini2000influence,rappa2016sperm}. Sperm
selection is also important for sperm cryopreservation. Established
selection techniques are moreover poorly effective in the most serious
cases of oligospermia ($< 4$ million/ml)~\cite{smith2017application}. Microfluidics is certainly a promising road for the
future of in-vitro fertilization~\cite{han2010integration,matsuura2013microfluidic,chen2013sperm,tasoglu2013exhaustion,huang2015fertilization,knowlton2015microfluidics,hussain2016sperm,zaferani2019strictures}.
Microfluidic
methods for sperm selection have been recently implemented~\cite{cho2003passively,schuster2003isolation}, demonstrating interesting capabilities and
important reduction of damaging probabilities~\cite{nosrati2014rapid,asghar2014selection,rappa2016sperm,shirota2016separation,xiao2021fertdish,simchi2021selection}. Many of these methods require the use of an external pumping in the
micro-flows through channels, which is a further source of possible
mechanical stress on the cells as well as a technological complication
affecting costs. Microfluidics-based techniques for sperm sorting without external flow pumping have also been investigated recently~\cite{nosrati2014rapid,xiao2021fertdish,simchi2021selection}. In these studies the largest effort is devoted to the design of a ready-to-use chip for biomedical applications and to the evaluation of  DNA fragmentation, while less attention is paid to 
 the role of geometry and trap shapes, looking for possible physical effects that can enhance the sorting capacity.
 
The improvement of automatic techniques for separating motile cells
from the non-motile ones can impact also diagnostic protocols. The
evaluation of sperm concentration is typically operated by direct
observation, under the optical microscope, in so-called Makler or
Neubauer cameras, with noticeable approximation. The assessment of
motility, not necessarily correlated with concentration, is another
factor of subjectivity~\cite{agarwal2016non} so that the classical male
fertility exam, the spermiogram, is considered to be of not really
high statistical significance~\cite{lu2010laboratory}. Less subjective approaches
include expensive systems such as CASA (Computer Assisted Sperm
Analysis) and flow cytometry~\cite{graham2001assessment}.  Recent studies have evidenced how microfluidics is a promising technology for sperm
diagnostics and point-of-care applications~\cite{zheng2011epetri}.

\begin{figure*}
 \centering
 \includegraphics[width=15cm]{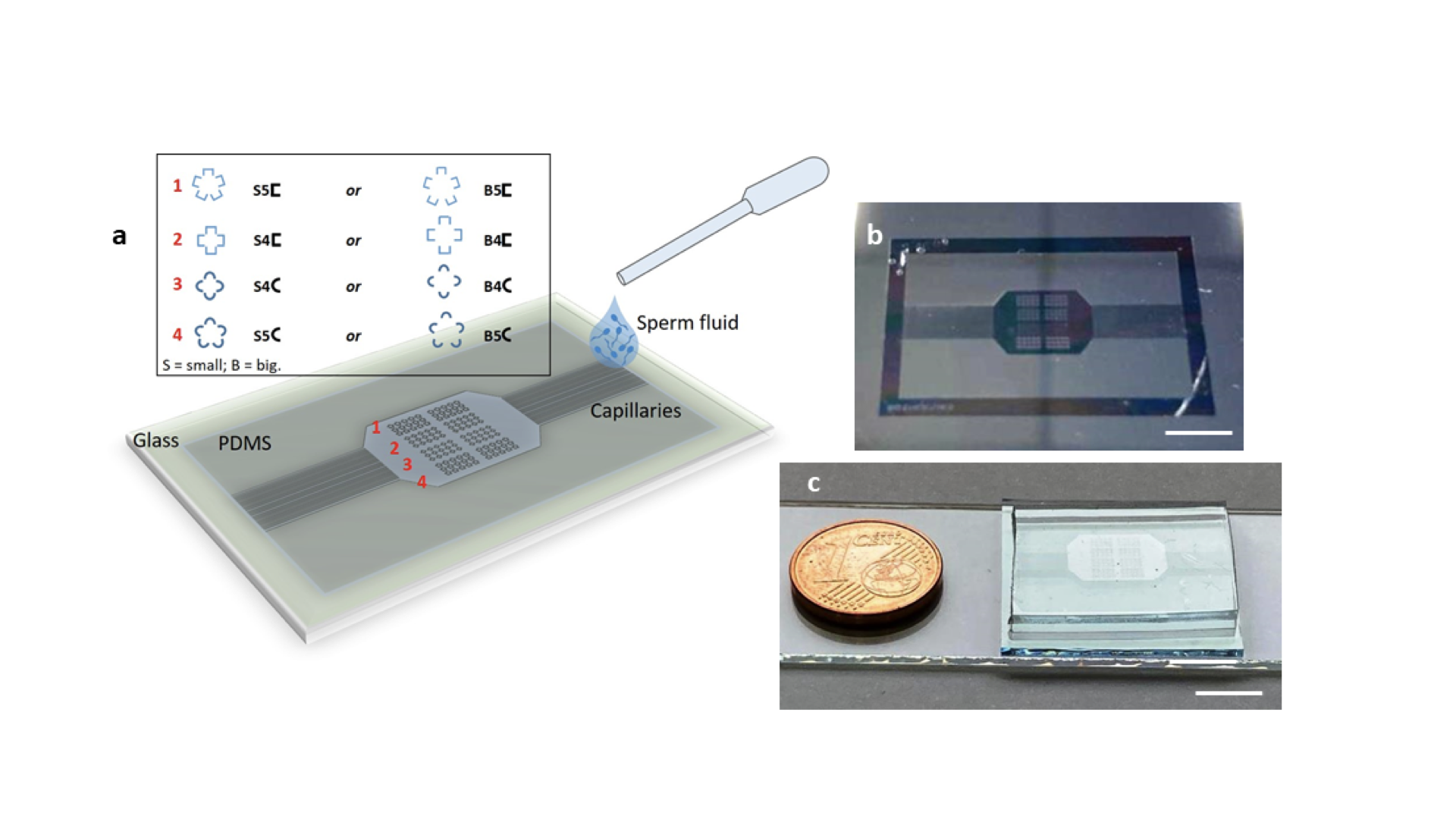} 
 \includegraphics[width=15cm]{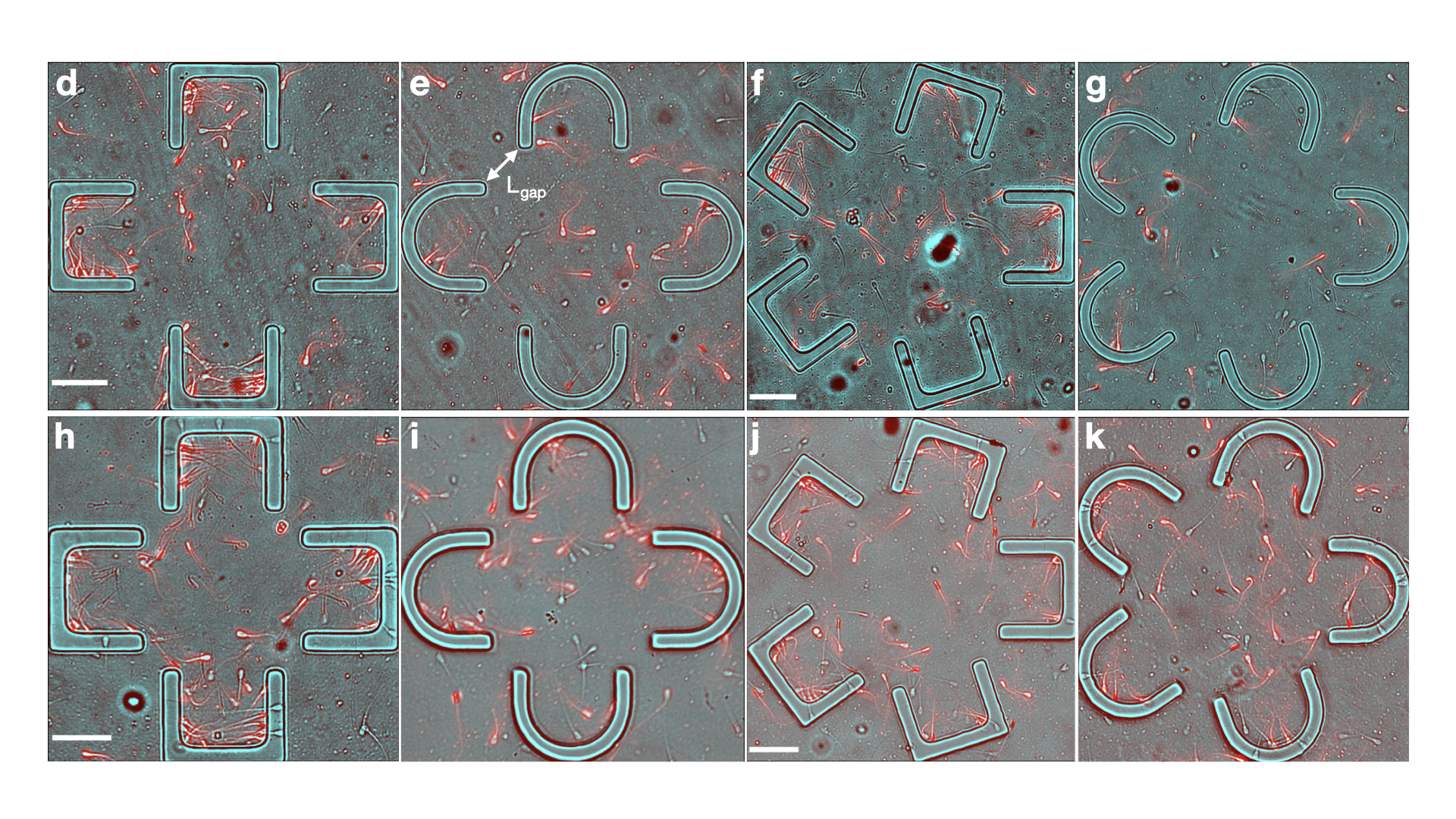} 
 \caption{{\bf Design of microfluidic devices. } a. Sketch of the quasi-2D hydrostastic microfluidic device showing the two networks of input /output capillaries connected to a cell sorter chamber including an array of microstructured trapping units. The chamber was filled, by capillarity, with bull sperm samples. The legend on the top shows geometric characteristics of the 4 groups of different shaped trapping units together with the corresponding acronyms used in the text. b. Picture of a patterned SU-8 master. c. Picture of the final microfluidic polydimethylsiloxane (PDMS) device for the passive trapping of sperms. Scale bars in b and c: $1 cm$.
(d-k): Microscope pictures with the design of the eight kinds of structures. The top row
  represents structures with big (B) gaps, $L_{gap} \approx 40 \mu m$, the lower row structures
  with small (S) gaps, $L_{gap} \approx 20 \mu m$. The pictures have been recorded in gray-scale from the
  microscope camera (see Methods) and coloured according
  to the following protocol. Red encodes motility, i.e. red level of
  each pixel indicates the variance of that pixel along a sequence of
  7 frames centred around a given frame. The blue and green levels
  put in evidence structures and non-motile sperm cells, as they just encode
  the light level of the original central frame. Scale bars in d,f,h,j: $50 \mu m$. }
 \label{fgr:setup}
\end{figure*}

Our aim here is to demonstrate a new class of sperm
sorting/concentration techniques which do not require external pumping in the micro-flows but
still take advantage of  low-cost and integrated microfluidic systems, obtained by soft-lithography microfabrication. This approach allows to enhance and modulate the confinement effects experienced by cells in their dynamics even in a restricted diffusion condition, typical of flow at the microscale.
In this case, a pumped active flow is not really necessary, because the cells are motile by themselves and a  separation can occur spontaneously. Indeed, non-motile (dead) sperms behave as passive
 particles dispersed in the fluid, and therefore diffuse very
slowly without preferences for particular regions of the channel.
The motile cells, as other kinds of cells equipped with self-propulsion (such as {\em Escherichia coli}, {\em Bacillus subtilis} and others)
 move very fast and display non-intuitive behavior in the presence of solid surfaces~\cite{elgeti2015physics,nosrati2015two,rode2019sperm}, particularly adhesion~\cite{rothschild1963non,li2008amplified,smith2009human,elgeti2010hydrodynamics,magdanz2015sperm}, swimming parallel to surfaces/obstacles~\cite{denissenko2012human,guidobaldi2014geometrical,nosrati2015two,rode2019sperm} but also long-time trapping in partially closed geometries~\cite{paoluzzi2020narrow,yaghoobi2021progressive}. 
The
consequence of this behavior for filtering purposes on general models
and on real bacteria have been studied theoretically and
experimentally~\cite{galajda2007wall,angelani2011active,koumakis2013targeted,koumakis2014directed}. From the point of view of statistical physics, the demixing (e.g. separation between motile and non-motile cells)
 originates from the contrast between opposite
thermodynamic situations, i.e. equilibrium (non-motile cells)
vs. non-equilibrium (motile cells). In the first case, one cannot
expect spontaneous de-mixing, as dictated by the second principle of
thermodynamics, whereas in the latter case de-mixing is not forbidden, and the spontaneous ratchet effect due to asymmetric geometries is a way to realise it~\cite{makse1997spontaneous,di2010bacterial,mccandlish2012spontaneous,gnoli2013brownian,medina2016cellular,chen2018chemotactic,striggow2020sperm}.

In our study we investigate experimentally a wide range of microstructures within an integrated microfluidic device
to determine    the key ingredients optimising the trapping/sorting power. A numerical simulation widen even more the
range of accessible structures, confirming and deepening our
understanding of the trapping mechanism. An important feature of sperm
passive trapping, already emerged in previous research~\cite{guidobaldi2014geometrical},
is the high trapping power of curved walls with small curvature
radius, such as corners~\cite{nosrati2016predominance}. Inspired by those previous works, we take advantage of this effect to increase the sorting capability of a micro-structured fluidic device. Thanks to this observation, the maximum relative concentration (ratio between the density of motile cells inside and outside of the traps) achieved in our study doubles what found using geometries with large curvatures.
We expect that our study can be a starting point for the development
of point-of-care applications for male fertility diagnostics and for
automatization of sperm sorting procedure finalised to the less
expensive in vitro fertilization techniques (e.g. IUI).

\section{Results: experimental}

\begin{figure*}
 \centering
 \includegraphics[width=18cm]{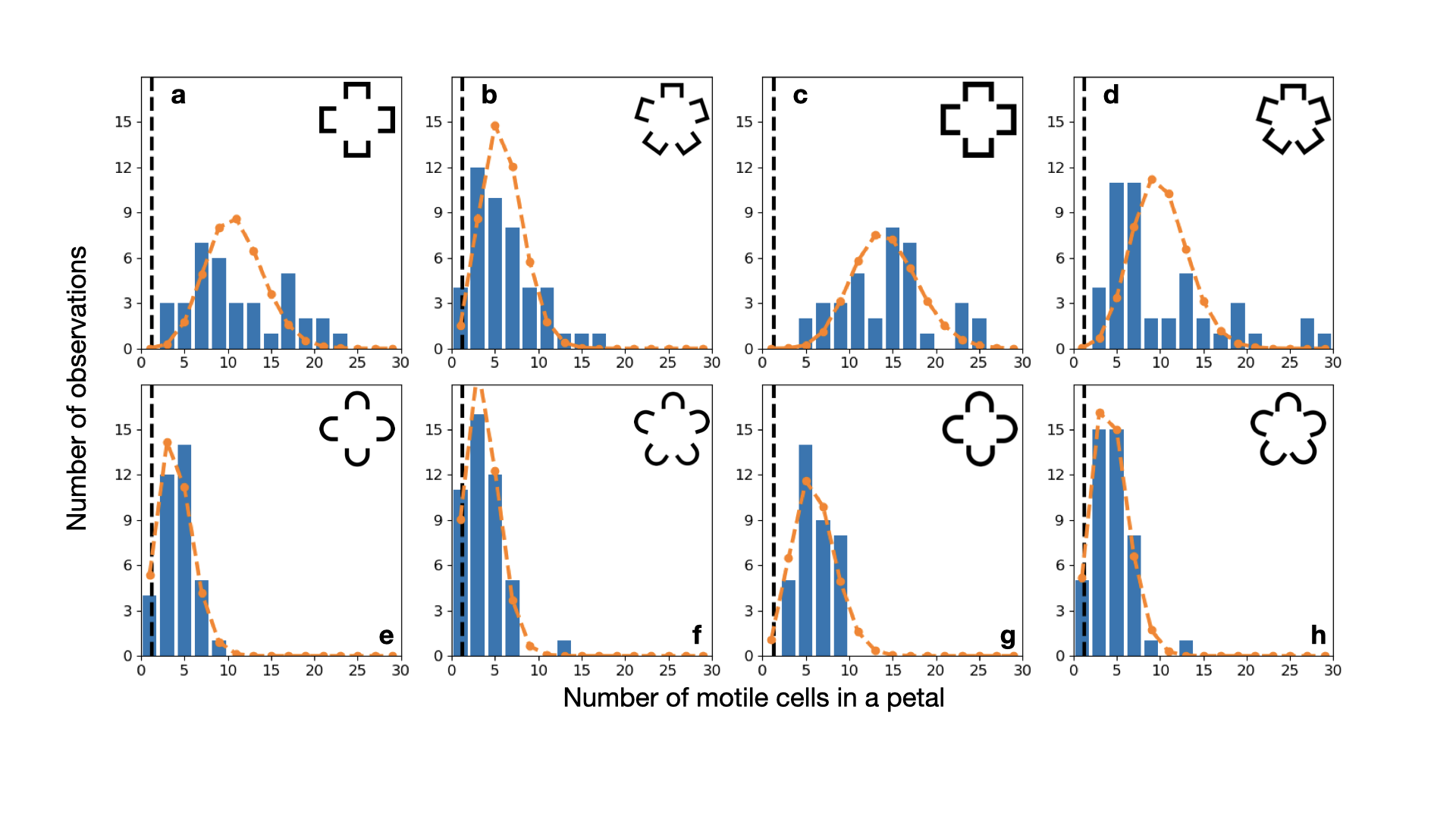}
\caption{
{\bf Concentration in petals (experiments).} 
Histograms of the observed numbers of motile cells in each petals. Each bin has  width $2$. The top four  histograms are for the structures with  
$\sqsubset$-shaped petals, 
the lower fourhistograms are for the structures with $\subset$-shaped 
petals. The first two columns refer to big gap microstructures, $L_{gap}=40\mu m$, while the last two are small gap microstructures, $L_{gap}=20 \mu m$.  It is  evident the larger trapping power of petals with $\sqsubset$-shape 
Poissonian distributions with the same average of the data are superimposed to the histograms in orange. Poissonian statistics is compatible with all data (p-value larger than $0.05$) apart from the first and last case of top row 
(B4$\sqsubset$ with $p=0.04$ and S5$\sqsubset$ with $p=10^{-5}$). 
The vertical dashed lines mark the number of motile cells that are found in the sperm liquid far from the structures, in an area equal to that of the petal. 
}
 \label{fgr:petals}
\end{figure*}

Trapping experiments have been carried on using quasi-2D hydrostatic microfluidic devices.  Each integrated device consists of two networks of input /output microchannels connected to the microstructured chamber (Fig. 1) and filled, by capillarity, with bull sperm samples after standard thawing protocols.
The microfluidic trapping devices are realized in polydimethylsiloxane (PDMS) by conventional soft-lithography and replica-molding technique, starting from patterned SU-8 master ad hoc fabricated by photolithography~\cite{xia1998}.  The chamber is designed as a functional cell sorter, with a total surface area of approximately $80\, mm^2$ and an internal volume of about  $1.2\,  \mu l$, and is micro-structured inside (Fig. 1a). A network of rectangular microfluidic-channels is used to allow the injection of the sample by capillary imbibition~\cite{viola2005} (see Methods for details).  PDMS was chosen to ensure optical transparency for the real-time microscopy imaging, compatibility with cells and biomolecules, flexibility and high conformability with the master structure (Fig. 1b)~\cite{zizzari2019}. The final sorting devices are produced by placing the microstructured polymeric replicas in conformal contact with glass substrates in order to obtain a perfect sealing of both microchannels and chamber. A picture of the final device is shown in Fig. 1c.
The dimensions and characteristics of both the microfluidic chamber and of microstructures guarantee spontaneous imbibition and capillary displacement of the sample without any external pressure. In a microfluidic device with very low aspect ratio microstructures, the fluid displacement is in fact mainly driven by restricted diffusion and laminar flow~\cite{viola2005,viola2014}.

Details for microlithography and sample treatment procedures are given in the Methods. The design of the functional chambers is based upon an array with 4 groups of different shaped trapping units, each group contains several identical units separated by an average period of about $450 \mu m$. Each trapping unit has a flower-like design made of n “petals” (with n = 4 or n = 5 and different shaped petals) around a central region: the spaces between petals are the inlets allowing sperm cells to enter or leave a trapping unit; in the following we call them “gaps”, their minimum distance is named $L_{gap}$ and in the two chamber designs it takes two values: $20 \mu m$ (“S” small) and $40 \mu m$ (“B” big). Each petal can have a “rounded” (\rsperm) or “cornered” (\csperm) shape, see Fig. 1a. A given structure is denoted by the combination of the three possible parameters $L_{gap} \times n \times \textrm{shape}$, e.g. B4\rsperm stands for 4 rounded petals separated by big gaps, and this makes a total of 8 different kinds of traps, each repeated for at least 9 times in order to collect a larger statistics in the results. Several experiments in re-printed channels with the same design (together with cumulated observations at different times) allowed us to increase even further the statistics. The average area of a trapping structure is $\sim 225\,\mu m\times
225\,\mu m$  and the height of the channel is $\sim 15 \mu m$, i.e. the trapping volume is - on average – $7 \times 10^{-7} ml$. At the typical observed sample concentration $1.2 - 1.3 \cdot 10^7$ motile cells per milliliter, one would find $\sim 8 -
9$ cells in a structure. On the contrary, the number of motile cells is found to be in the range of  $20 - 80$, corresponding to a density enhancement of a factor $3 - 10$, depending upon the choice of $n$, $L_{gap}$ and the shape of the petals, as detailed in the rest of the work.

\begin{figure*}
 \centering
 \includegraphics[width=20cm]{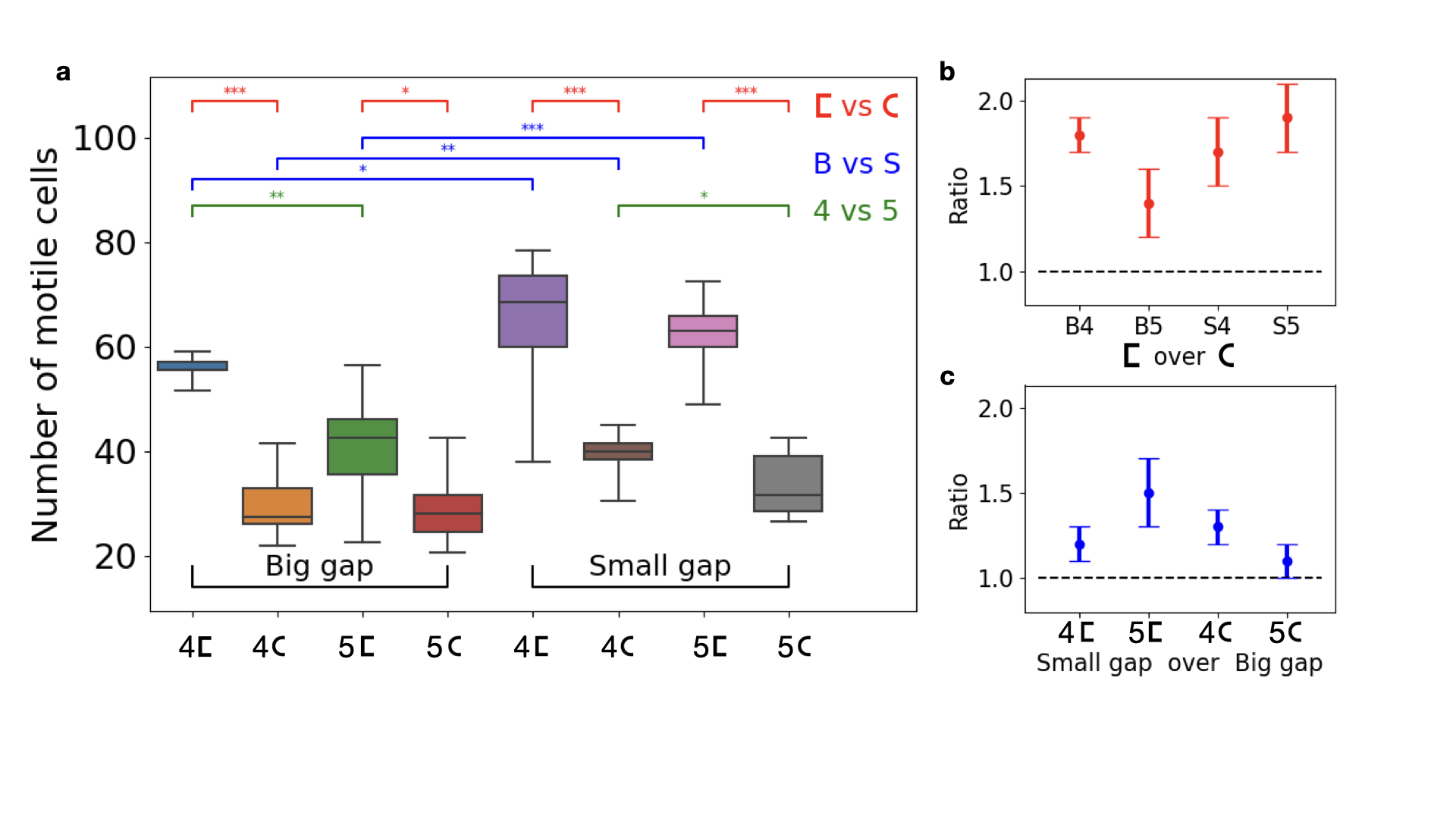} 
 \caption{{\bf Concentration in structures (experiments).} Total number of motile cells in each structure (experiments). a. Box-plot representing the distributions over several structures of the same kind:  each colored box contains a line representing the median, its size covers the quartiles of the dataset, while the whiskers extend to show the rest of the distribution. On the top of the graph we have included the results of statistical t-tests comparing pairs of distributions: each segment connects two significantly distinct distributions, with $p$-value smaller than $0.05$ (*), $0.005$ (**) and $0.0005$ (***). Pairs without a connecting segment are not significantly distinct.  b. Ratio of averages numbers between the case of $\sqsubset$-shaped petals and $\subset$-shaped petals, keeping fixed $L_{gap}$ and the number of petals. c. Ratio of averages numbers between the case of small $L_{gap}$ (“S”) and big $L_{gap}$ (“B”), keeping fixed the  petal shape and the number of petals. In plots  b and c  the error bars represent the propagation of standard deviations. }
 \label{fgr:total}
\end{figure*}

One of the main results of our work is the role, in the sperm trapping mechanism, of petals, particularly of their shape.
Histograms of petal occupation, shown in Fig. 2, announce the results of
the study of whole structures (below).  The averages of each distribution is much larger than the average number of cells that is found in an equivalent area outside (and far from) the traps, marked by the black dashed vertical line in each graph. It is also evident that more cells are
captured by \csperm-shaped petals rather than \rsperm-shaped ones, by smaller gaps (“S”)
rather than large ones (“G”), while the number $n$ of petals seems to have
a less clear influence on cell trapping.
Distributions are compatible with Poissonian statistics
(p-value larger than $0.05$) apart from two cases (B4\csperm with $p=0.04$
and S5\csperm with $p=10^{-5}$). It is tempting to deduce that there is no
correlation induced by interactions between the cells, however
direct observation suggests that the dynamics - particularly the
movement from one petal to another - often occurs in groups of few
coordinated cells (see Movies in the ESI). Indeed dynamical correlations are not incompatible
with Poissonian steady state statistics, but they are difficult to be
measured.

In Figure 3a we show results about the occupancy of the whole
structure, which means counting the sum of motile cells in all the
petals and in the central region of each structure. See the Methods
section for details about the way the statistics is collected. The
plot teaches us with fair accuracy that the number of motile cells per
structure is affected positively by the shape of the petals (\rsperm-shape is surpassed by
\csperm-shape), and by the size of the gap (structures with big gaps are surpassed by structures with small gaps). These trends are quantified by statistical t-tests over couple of distributions, using *s to mark how significant is the difference~\footnote{Our observations of Fig. 2 indicate a compatibility of cell count fluctuations with Gaussian statistics, justifying the uses of t-test. We have also confirmed our results through a non-parametric Mann-Whitney test.}. Higher significance is observed when the shape is changed.
In Figure 3b we show 
the ratio between the average numbers of motile cells when one parameter is changed and the others are not, considering only changes in shape and in $L_{gap}$,. It is seen that cornered structures, with respect to rounded ones,
increases the number of trapped motile cells by a factor significantly
larger than $1$, up to a factor $2$ in the case of small gaps with $5$
petals.

\begin{figure*}
 \centering
 \includegraphics[width=18cm]{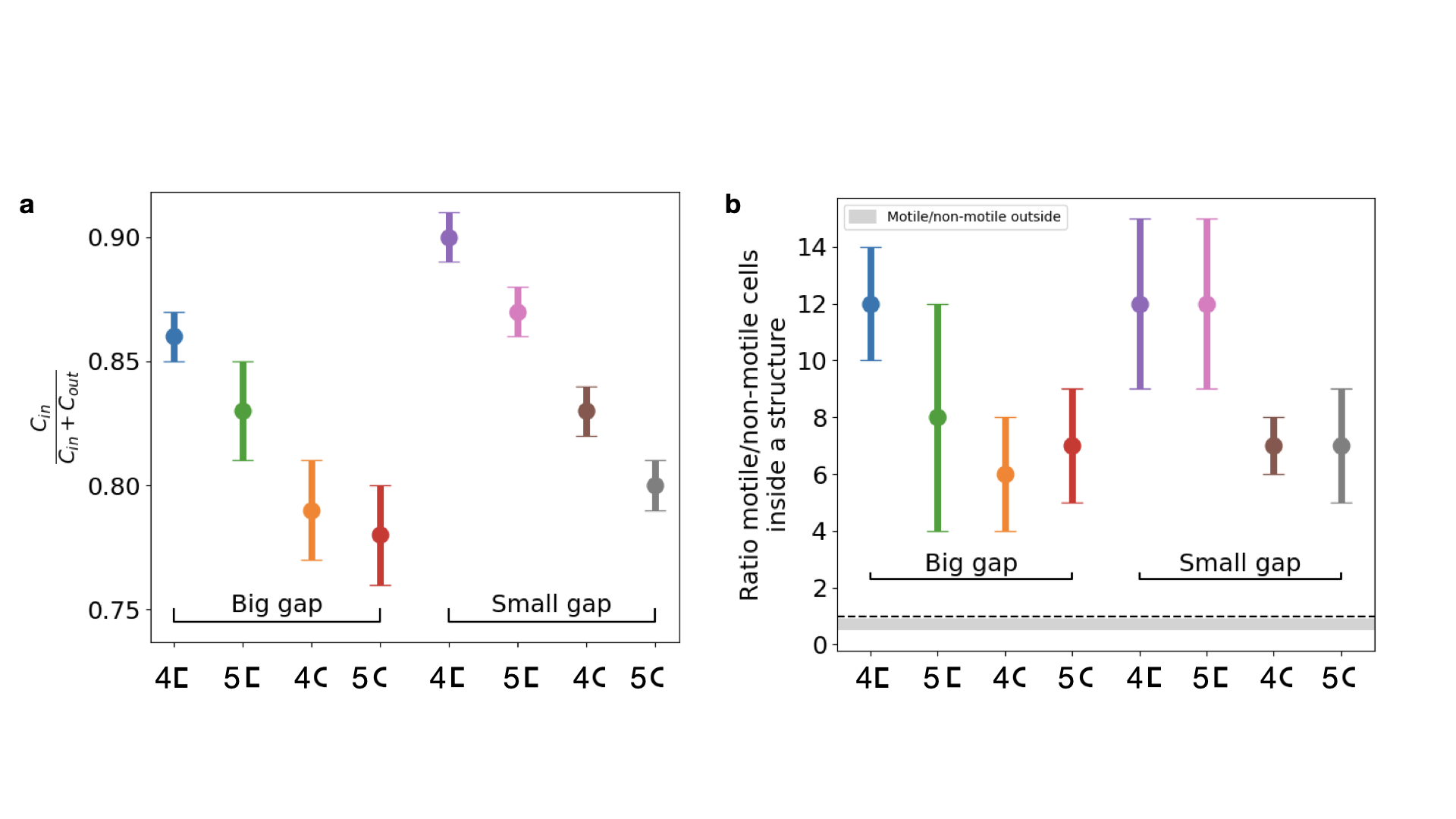} 
 \caption{{\bf Relative concentrations (experiments)}. a. Relative concentration (averaged over 9 structures of the same kind)  of motile cells between inside and outside of a structure. Error bars are standard deviations. b. Ratio (averaged over 9 structures of the same kind)  between motile cells concentration and non-motile cells concentration, always inside a structure. Error bars are standard deviations. For reference the gray area represents the same ratio measured outside of the structures. The width of the gray area is the standard deviation.  The dashed line marks equal concentrations (ratio  $1$).}
 \label{fgr:relative}
\end{figure*}

In Figure 4a we test our previous conclusions against the variability of the area of the structures as well as 
of (motile) sperm cell concentration in the different samples and in
the different regions of the same channel. We need to rule out that
results for the numbers of cells in the structure are not 
influenced by fluctuations of cells in the region and sample where we
have performed the observation, which is known to display some (weak) variation. Plot 4a illustrates the relative concentration 
of motile cells $C_r=C_{in}/(C_{in}+C_{out})$: $C_{in}$ is the average concentration (number per unit area) of cells inside a structure, $C_{out}$ is the average concentration outside of the
the structure, measured roughly in the mid distance between the structure and nearby
structures (this ensures that surface effects are minimised). The relative concentration is defined to be constrained between $0$ (no
trapping) to $1$ (infinite trapping), while the concentration ratio $C_{in}/C_{out}=C_r/(1-C_r)$  can grow to infinite. We remark that, with cornered structures and small gaps, we
get quite high values of this ratio: $C_{in}/C_{out}\sim 7 - 9$. The use of cornered structure is the key to get a concentration contrast much higher than the one  achieved in previous studies~\cite{guidobaldi2014geometrical}. The role of the curvature of walls is discussed in details in Section 3 and in the ESI where our numerical results show that, at constant physical dimensions, smaller curvatures always enhance the trapping power of structures or membranes. The effect of the number of petals $n$ going from $4$ to $5$ on the relative concentration is a reduction, of roughly a factor $\approx 1.2$ for all cases. This reduction is compatible, quantitatively, with the reduction, when going from $n=4 \to n=5$, of the percentage of area occupied by petals - which have the largest trapping power - with  respect to the whole structure.  

The last  observation we report is crucial for applications, as it concerns the ratio among
motile and non-motile cells - details on their distinction are contained in the Methods section "Distinction between motile and non-motile cells". The efficiency of a sperm sorting technique, in fact, is determined by the concentration of motile cells. The motile/non-motile ratio, if measured outside of the
structures, in the free area not too close to the outer boundaries of
the trapping structures, is quite unvaried in all the observed
samples, and amounts to $\sim 0.7 \pm 0.2$, i.e. in the non-trapped
regions the sample hosts a majority of non-motile cells. On the contrary the
ratio inside the trapping structures is clearly in favour of the
motile cells, see Figure 4b, with values in the range $4 - 15$,
depending on the structure type. The larger motile/non-motile ratios are
measured for cornered structures with small gaps. The ratio between motile and non-motile cells has larger errors with respect to $C_{in}/C_{out}$. 
The percentage of motile cells is much higher than $1$ for all the structures, the largest value are reached in those with cornered petals. 
This measure suggests that our mechanism acts as an efficient selectors for the motility of sperm. 
This result is crucial in applications that usually require not only high concentrations but also cells with high motility.


\section{Results: numerical}

We have adopted a simplified model of the kind of “wagging” Active
Brownian particles, inspired to previous studies~\cite{tasoglu2013exhaustion,guidobaldi2014geometrical}. In the
Methods we give the details of the model, here we describe it
qualitatively in its essential aspects. Each cell is represented by
its center of mass position and by the orientation of the
self-propulsion direction (e.g. the tail average direction) in the
plane. The orientation vector diffuses according to rotational thermal
Brownian motion with a very long persistence time (unperturbed sperm
cells follow straight paths longer than any characteristic length in
our experiment). The cells interact with each others through a soft
potential. The walls exert a harmonic elastic repulsion together with
a torque that tends to align the tail orientation to a direction
forming with the wall a small angle  (this is in accordance with
“occupied cone” models that reproduce the effective excluded volume of
the long tail~\cite{elgeti2010hydrodynamics,denissenko2012human}) . The “wagging” ingredient consists in an oscillating
force perpendicular to the cell orientation with a frequency
comparable to that of sperm beating~\cite{tasoglu2013exhaustion,gong2021reconstruction}. We have verified that all such
ingredients are necessary to obtain a behavior which is coherent with
our experimental observations. In particular the transversal
oscillation is key to obtain a trapping power of corners which depends
upon their curvature.

\begin{figure*}
 \centering
 \includegraphics[width=18cm]{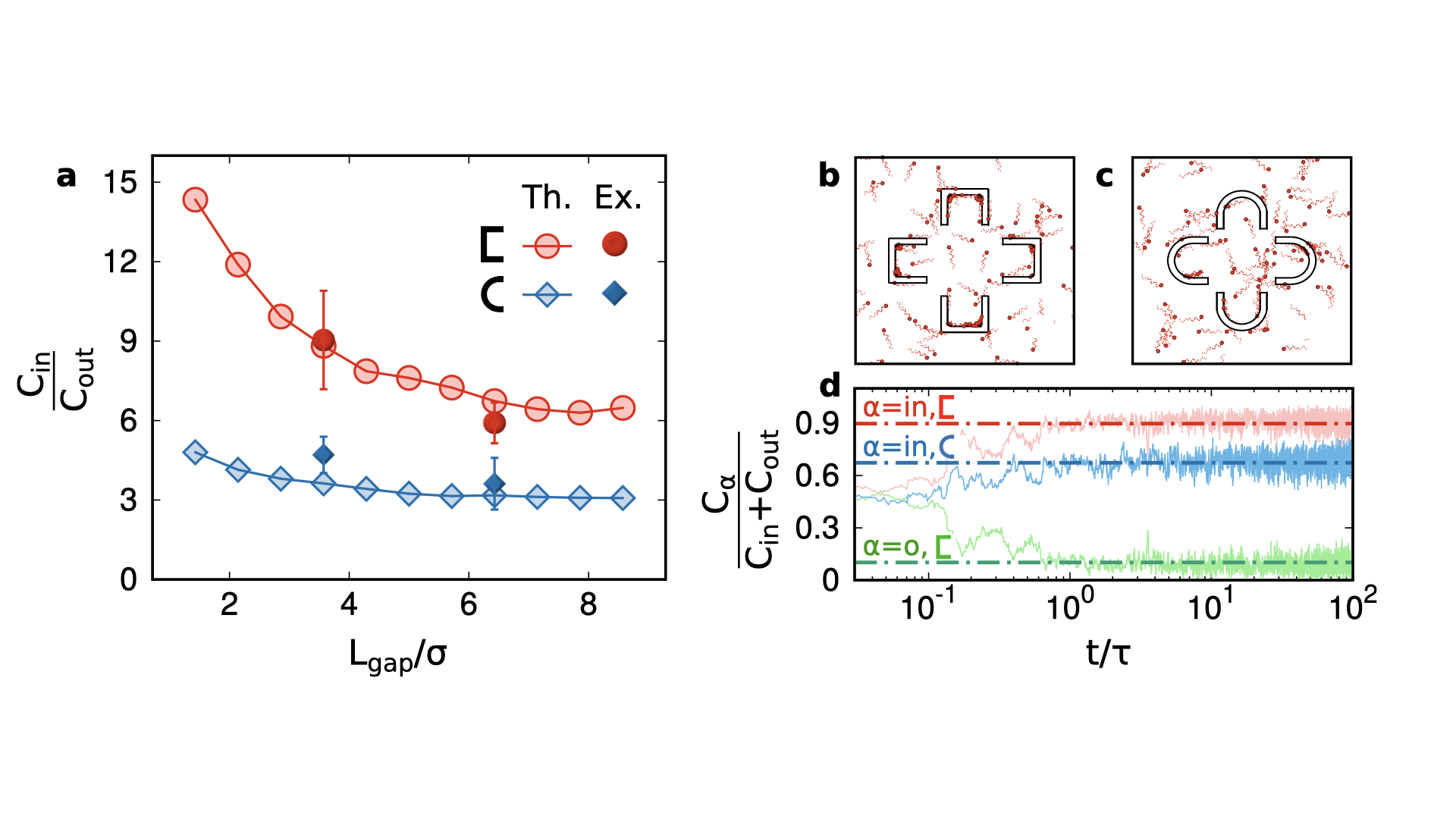} 
 \caption{{\bf Results from simulations.} a. Ratio between the average concentration of cells found inside and outside of a structure versus the width of the gap $L_{gap}$ (normalised by the sperm head diameter $\sigma$, see Methods); b-c. Snapshots of simulations with different structure shapes; the snake-like lines following each sperm position are obtained by tracing its motion in a short period of the transversal force oscillation (see Methods). d. Relative occupancy as a function of time, confirming that the initial concentrations are equal and then change in time, the outside concentration is slightly depleted while the  inside one increases  to  a  plateau  which depends upon the petal shape.}
 \label{fgr:simul}
\end{figure*}

The sizes of the structures, as well as the intrinsic properties of the sperms, such as speed, wagging frequency and persistence time, are fixed by  experimental observations.
The other parameters, such as the constants appearing in the repulsive potentials (for sperm-wall and sperm-sperm interactions) as well as constants in the aligning torque terms (see Methods), are fixed looking for reasonable comparison
between numerical simulations and experiments for a single choice of $L_{gap}$ and shape of the petals. After the calibration of the
parameters we have performed simulations with different $L_{gap}$ and different shapes of the petals, and
checked that the comparison with the experiments remained good,
obtaining a fair check of our assumptions (see Figure 5b,c for
snapshots, see also movies attached).

The main result of our simulations is the trapping power for two kinds
of pocket shape (cornered and rounded) and a wide range of gap sizes,
Figure 5a. The figure demonstrates that cornered \csperm structures are always more trapping than rounded \rsperm structures   and that increasing the gap size reduces the trapping
power. These two observations are in substantial agreement with the experiments. 
Remarkably, numerical simulations teach us that $L_{gap}$ is certainly a less relevant parameter than the shape of the petals. Indeed, the red curve in Fig.~5a is always above the blue curve, that is with \rsperm structures - whatever small is the gap size - one cannot achieve the relative concentration obtained with \csperm structures. 
We recall that decreasing $L_{gap}$ at constant size of the petals implies a relative increase of the trapping surface: this is the qualitative explanation of the decreasing curves in Fig. 5a.

It is interesting to observe that in both cases an asymptotic
trapping power larger than $1$ is found
suggesting that even spare petals (not close in “flowers”) can be used to entrap sperms. Of course increasing much more $L_{gap}$ would result in totally open structures and the measurement, in an experiment with a single structure surrounded by infinite unconfined volume,  would be dominated by bulk concentration, therefore a decrease to $1$ is expected, asymptotically: however verifying this requires much larger simulation chambers to avoid finite-size effects. We remark that such finite-size effects are present also in our experiments. It would be interesting, in future experiments, to study if the  relative position or orientation of the structures with respect to the entrance  can influence the filling of each structure by capillary injection. We speculate that a more dense arrangement of traps (i.e. drastically reducing the distance between them) could lead to a reduction of the trapping performance, as the original sample,  being loaded from a side of the chip,  would find more difficult to reach all the trapping units. 
Panel d of Fig.~5 is useful to evaluate the sorting time. The relative concentration, starting from a situation $C_{in}=C_{out}$, achieves an unbalanced plateau in a time $t \sim  \tau$ where $\tau$ is the persistence time of the model sperm cells (see  Methods). In physical time this corresponds to $\sim 100$ seconds. This result gives also a hint about the fact that the experimental measurements, taken a few minutes after the chip filling, come from a steady regime.

Numerical simulations (not shown) with single sperm cells in a pocket with different  curvatures radii $R$ confirm the origin, discussed already in~\cite{guidobaldi2014geometrical}, that trapping is enhanced by radii smaller than the occupied size of the sperm cell, typically head diameter $\sigma$  plus oscillation amplitude $A$, e.g. curvature with  $R \ll \sigma+A$ are strongly trapping with respect to curvature with  $R \gg \sigma+A$: this can be understood as a consequence of the fact that
 the interaction with a curved wall together with the
transversal oscillation  make very unlikely, for a sperm, to
follow the curvature without bouncing back. On the contrary, a sperm
model that has no wagging, can align close to and follow any
kind of curvature without never coming back, emerging from the pocket and
escaping from it. We refer to the Supplemental Movies for a rapid understanding of this scenario.

\section{Conclusions}

Motile cells, such as sperms or bacteria, display fascinating behavior which defies several basic aspects of the physics of mixtures realised with colloids, emulsions, suspensions etc. Segregation of different parts of a standard mixture (one made of components which are not self-motile) is usually  achieved by driving the mixture through externally imposed flows (such as centrifugation at the macroscale, or pumping a flow in microfluidic devices). The same happens, in standard  sorting techniques, for sperm samples~\cite{boomsma2004semen,rappa2016sperm}. 
In this work we have taken advantage of the intrinsic motility of sperms to obtain the segregation of motile cells in small confined regions. We have explored a family of microfluidic devices where three different parameters are tuned experimentally, in order to understand which is the optimal strategy to enhance the concentration of motile sperm. We have also calibrated a simple but effective numerical model of sperm cells (including interactions between cells and between each cell and the boundaries) which reproduces the experimental results and open a virtually infinite set of future explorations for improving the designs towards other needs and applications. We have put in evidence - experimentally and theoretically - that the efficiency of our sorting method, is mainly enhanced by the specific design of surfaces. Counterintuitively, their shape, in particular, small curvatures and cornered angles, plays a more important role than the size of the entrance (gap) of the trap: our device increases the concentration of motile sperms by an order of magnitude.
This has been explained in terms of the increased  self-trapping mechanism caused by the interplay between cornered angles and the the peculiar swimming mechanism of sperms.
The recent developments in  microfluidic designing techniques offer a  range of intriguing possibilities to further improve the efficiency of our trapping mechanism, for instance by manipulating the roughness of surfaces~\cite{guidobaldi2015disrupting}, creating a plethora of additional angles with small curvatures.

Recent studies have already exploited microfluidics without flow-pumping for sperm sorting~\cite{nosrati2014rapid,xiao2021fertdish,simchi2021selection}. These works are somehow complementary to ours, because they focus on technology readiness of the chip and on the evalutation of DNA fragmentation, while our analysis explores the geometry of the device, looking for optimal shapes that can enhance physical effects for sorting. Moreover, the filtered sample obtained within our chip (that is the content of the trapping chambers) has a cell concentration which is higher than the original sample, that is our chip is not only a filter/sorter but also a concentrator. Our device improves also the time-efficiency of the sorting process: in our study, the sorting time is of the order of a few tens of seconds, which is smaller than the typical times needed in previous studies~\cite{nosrati2014rapid,xiao2021fertdish,simchi2021selection}.  The device proposed here is low-cost, disposable if needed, easy to fabricate and use, and suitable for later integration with more complex devices for clinical applications. Moreover, it is at quite a high level of technological readiness. 
In the future, such a trapping device may be implemented in a multiplexed system for the purpose of picking up highly motile cells. The motile sperm cells, trapped in the microstructures, can be taken away with subsequent rinses or by providing apertures on a side of the chip and using a microscope-guided syringe to collect the concentrated sperm.
For diagnostics purposes, one may imagine optical reading  (through lighting levels) of the concentration in the confining region. A future development of our study concerns the correlation between sperm performance, e.g. speed or other parameters, and the features of the confining regions, which could help in designing more sophisticated chips for point-of-care diagnostic applications where a detailed spermiogram can be directly elaborated.

Our study suggests conjectural applications where oocytes are directly inserted into the trapping chambers of the chip of a microfluidic multilayered open device~\cite{oh2017open}: the diameter of mature oocytes, including the zona pellucida,  is smaller than 150-120 $\mu m$, comparable to the diameter of the central part of our chambers, which - however - can be easily redesigned with a slightly larger internal area, if necessary. This idea could be interesting for future researches, and could also lead to a totally new in-vitro fertilisation technique, in principle more effective than conventional IVF (thanks to the high concentration of motile, already selected, sperms around the oocyte) and where the final selection of the fertilising sperm is not due to the hand of a technician/biologist, as in ICSI, but to natural competition in oocyte penetration among several tens of sperms. A further advantage of such a conjectural technique is that semen needs neither preparation nor waiting in the dish/chip for the oocytes: as soon as the oocytes are ready (e.g. after incubation and other treatments  which usually take hours) they are transferred in the chip trapping chambers and only then the raw sample, just after ejaculation or thawing, is injected by capillarity in the chip: at that point in a few seconds the chambers  fill up with dense concentrations of $\sim 90\%$-motile sperms, in close contact with their target.

\newpage
\section{Methods}

\subsection{Lithography of the structures}

Microfluidic trapping devices are fabricated using conventional soft lithography and replica molding technique from SU-8 (Microchem, YMC, Switzerland) masters.
SU-8 masters are realized via standard optical lithography~\cite{zizzari2011,chiriaco2016}: a thin layer of SU-8 negative photoresist is deposited onto a clean silicon substrate and the entire patterns of ad hoc designed photomasks (J.D. Phototools Ltd., Oldham, Lancashire, UK) are transferred on the photoresist film via UV exposure. The process has been optimized to obtain a height of ($14.7 \pm 0.5) \mu m$ for all the characteristic structures of the device.
Each device consists of two networks of input /output microchannels connected to a microstructured chamber (Fig. 1a) in order to fullfil sperm samples by capillarity. Rectangular microfluidic channels have dimensions of $100 \mu m \times 15 \mu m$  and length of about $10 mm$.  The quasi-2D microstructured chamber has total dimension of $10 mm \times 8 mm \times 15 \mu m$  and includes a $12 \times 10$ matrix with 4 groups of different trapping units: each group contains at least 30 identical trapping units separated by an average period of about $450 \mu m$. As shown in Fig. 1a the trapping area consists of a flower structure characterized by: i) a different number of “petals”, n = 4 or 5; ii) different shape of petals, rounded (R) or cornered (C); iii) different opening spaces at the base of each petal, “gap” $L_{gap}$, which are small (S, with $L_{gap}= 20 \mu m$) or big (B, with $L_{gap}= 40 \mu m$). The combination of these features creates a heterogeneous matrix for trapping analysis. The different units alternate every 3 rows of the matrix in order to ensure that they are all uniformly reached by the sperm cells via capillary motion and to collect a larger statistics in the results.
Thereafter, the microfluidic trapping chips are obtained by replica molding, by casting a mixture of PDMS pre-polymer and curing agent (10: 1, Sylgard-184, Dow Corning - USA) onto one of the SU-8 masters. The PDMS is then polymerized at 140° C for 15 min and then detached from the master. To promote the imbibition, the microchannels are opened with a razor blade and the replica is placed in conformal contact with a glass substrate to allow the closure of the device. To improve sealing and displacement of cell sample, both the replica and the glass are treated with oxygen plasma (Diener Pico, low pressure plasma system: $100 W$, $240 cm^3/min$ of $O_2$ flow; $0.5 mBar$; $60 s$). 

\subsection{Details about sperm solution preparation}

Spermatozoa samples from bulls were obtained from "Agrilinea S.R.L" Rome, and preserved in a liquid nitrogen cylinder. The sperm samples were obtained in the form of sperms suspended in semen and packed in several vials. At the onset of the experiment, one vial of sperms was taken out of the nitrogen cylinder and immersed in a hot water bath of 37 $^\circ$C for 10 minutes. After 10 minutes, the vial was taken out of the bath and immediately cut open using a pair of sterilized scissors and the entire content was poured out of the vial into an Eppendorf. Using a micropipette,  50$\,\mu l$ of the sperm suspension was sucked out from the Eppendorf and inserted into the microchannel, ensuring proper filling inside the structures. The ends of the channel were then sealed. The locomotion of the sperm cells were recorded by using a digital camera (Nikon, USA) connected to an inverted microscope. During the entire duration of the experiment, the microchannel placed on the stage of the microscope was subject to a controlled temperature environment of 37 degrees centigrade using heating unit and temperature sensors. An in house software developed using Python Programming language was used to capture and analyse the images. Naked-eye cell counting has been performed on the image sequences (movies  $20$ seconds long with $50$ frames per second acquisition), where each image has a $2048 \times 2048$ pixels resolution and captures (at magnification $10x$) an area of $\sim 1330 \times 1330 \mu m^2$ (a matrix of $3 \times 3$ structures). Counts have been repeated every $5$ seconds to ensure stability of the numbers. The occupancy of each petal or structure is defined as the average between the count in the first $5$ seconds and the count in the lat $5$ seconds. We waited $\sim 5-10$ minutes, after the chip filling, to start image acquisition: this (according also to our numerical simulations, see Fig. 4d) guarantees a stationary state.

\subsection{Distinction between motile and non-motile cells}

In our experimental study we have distinguished the class of motile and non-motile cells by means of a threshold on their velocity, at $1  \mu m / s$. The choice of this value comes naturally from empirical evidence and is not particularly significant within a broad interval. Indeed, all our experimental observations come from $20$ seconds-long recordings, and we have a resolution of slightly less than a micron per pixel. As a consequence, when we label a cell "non-motile", we can conclude that it has moved less than $1$ micron in $20$ seconds, i.e. it has a speed smaller than $0.05 \mu m/s$. Cells with a velocity larger than this threshold have a non-negligible velocity that can be measured with good resolution and that allows us to label a cell "motile". However, we have rarely observed cells slower than $10 \mu  m/s$, so that motile and non-motile cells are well-dinstiguished.



\subsection{Model for the simulations}

To reproduce our experimental findings, we model each sperm as a spherical active particle in the overdamped regime. 
The complex swimming mechanism of a single sperm is reproduced by effective time-dependent forces \cite{guidobaldi2014geometrical}, included in the dynamics of the center of mass position, $\mathbf{x}_i$:
\begin{equation}
\label{eq:Method_motion}
\gamma\dot{\mathbf{x}}_i=\mathbf{F}_i + \mathbf{F}_i^w + \gamma v_0 \mathbf{n}_i  + \sqrt{2 \gamma  T} \,\boldsymbol{\eta}_i + A\omega \cos{(\omega t)}\mathbf{n}^{\perp}_i\,,
\end{equation}
where the constant $\gamma$ is the drag coefficient and $T$ the solvent temperature.
The term $\boldsymbol{\eta}_i$ is a white noise vector with zero average and unit variance accounting for the collisions between the solvent molecules and the active particles, such that $\langle \boldsymbol{\eta}_i(t) \boldsymbol{\eta}_j(t')\rangle=\boldsymbol{\delta}(t-t')\delta_{ij}$.

The particles interact through the force $\mathbf{F}_i = - \nabla_i U_{tot}$, where $U_{tot} = \sum_{i<j} U(|\mathbf{x}_i -\mathbf{x}_j|)$ is a pairwise potential. 
The shape $U$ is chosen as a shifted and truncated Lennard-Jones potential:
\begin{equation}
\label{eq:interactionpotential}
U(r) = 4\epsilon \left[ \left(\frac{\sigma}{r}\right)^{12}- \left(\frac{\sigma}{r}\right)^6  \right] \,,
\end{equation}
for $r\leq 2^{1/6}\sigma$ and zero otherwise.
The constants $\epsilon$ and $\sigma$ determine the energy unit and the nominal particle diameter, respectively.
The term $\mathbf{F}_i^w$ represents the repulsive force exerted by the obstacles, whose properties will be specified later. 

The effects of the flagella are modeled by the active force, $\gamma v_0 \mathbf{n}_i$, evolving according to the active Brownian particles (ABP) dynamics.
In the ABP model, the active force acts locally on each particle, providing a constant swim velocity $v_0$ and a time-dependent orientation, $\mathbf{n}_i=(\cos{\theta_i}, \sin{\theta_i})$. The orientational angle $\theta_i$ evolves stochastically via a Brownian motion:
\begin{equation}
\label{eq:theta}
\dot{\theta}_i = \sqrt{2D_r} \chi_i + T_i^w\,,
\end{equation}
where $\chi_i$ is a white noise with zero average and unit variance and $D_r=1/\tau$ determines the persistence time of the active force. $T_i^w$ is the torque exerted by the wall whose properties will be explained in detail later.

The last term in the right hand side of Eq.~\eqref{eq:Method_motion} is a periodic time-dependent force which mimics the oscillations of the sperm head because the vector $\mathbf{n}^{\perp}=(- \sin{\theta}, \cos{\theta})$ points perpendicularly to the swimming direction $\mathbf{n}_i$.
The constants $A$ and $\omega = 2\pi \nu_i$ determine the amplitude and the frequency of the oscillations, respectively.

\subsubsection{Force and torque exerted by the walls}

Each wall is described by a continuous closed line in the plane, representing its perimeter, which in some parts is external to the structure and in some other parts is internal. Locally (near the point of contact) it can always be written as $y=w(x)$ and/or $x=w(y)$. Let us consider one of the two cases (the first), while the other is obtained by simply exchanging $x$ with $y$.  In the first case the force exerted by the wall reads:
\begin{equation}
\mathbf{F}^w= - U'(w(x)-y) \mathbf{e} \,,
\end{equation}
where $\mathbf{e}=(w'(x), -1)/\sqrt{1+w'(x)^2}$ is the unit vector orthogonal to the wall profile and
$U_w(r)$ is chosen as a harmonic repulsive potential truncated in its minimum:
$$
U_w(r)=\frac{K}{2} r^2 \Theta(r) \,,
$$
where $K$ determines the strength of the potential and $\Theta(r)$ is the Heaviside step-function.

The torque exerted by the walls, $T^w_i$, reads:
\begin{equation}
\begin{aligned}
T^w_i({\theta, r, \phi})=& - I(\theta,\phi) \,\ell_0 |\mathbf{F}_w| \sin[2(\theta- \phi \pm \alpha)] \,
\end{aligned}
\end{equation}
where the angle $\phi$ defines the orientation of the wall and is defined as
$$
\phi=\psi+\frac{\pi}{2}
$$
being $\psi$ the angle formed by the outward normal with respect to the $\hat{\mathbf{x}}$ axis.
The function $I$ selects the interval of $\theta$ which activates the torque, namely when the self-propulsion vector points inward with respect to the wall profile. The torque aligns the particle orientation to the orientation of the wall so that an angle $\pm\alpha$ (the sign is determined by the orientation) is formed. This assumption is consistent with experimental results since each sperm usually does not swim parallel to the wall but forms a relative angle of $\approx 10$ degree~\cite{elgeti2010hydrodynamics,denissenko2012human,bettera2020hitting}. We remark that the torque ingredient is fundamental to avoid accumulation of particles (without sliding) close to the wall, as it is observed in simulations without torque~\cite{caprini2020activity}.

\subsubsection{Simulation parameters \& geometrical setup}

The typical size of the particle is chosen as $\sigma=6 \mu m$, and accounts for the effective volume occupied by the head of each sperm, which is approximated as a disk.
The persistence time is $\tau=10^2 s$, while the swim velocity $v_0=30 \mu m/s$.
The parameters of the thermal bath, $\gamma$ and $T$ gives rise to a diffusion coefficient $D_t=10^2\mu m^2/s$ much smaller than the effective diffusivity due to the active force, $D_a=v^2_0 \tau=9 \times 10^4 \mu m^2 /s$, as it usually happens for self-propelled particles~\cite{bechinger2016active}.
Finally, amplitude and frequency of the head oscillation read $A=3.5 \mu m$ and $\nu=30/(2\pi) s^{-1}$, respectively, in agreement with Ref.~\cite{guidobaldi2014geometrical}.
The energy scale of the interaction is chosen as $\epsilon=1 \mu m^2/ s^2$, while the parameters of force and torque of each wall are $K=10^3 \mu/s^2$, $\ell_0=6 \mu m^2 / s^2$ and $\alpha=10$ degree.

Simulations are performed using dimensionless units, rescaling time and position through the time scale introduced by the active force and size of the sperm, namely $\bar{t}=t/\tau$ and $\bar{x}=x/\sigma$.

The numerical study is performed by simulating a system of $N=100$ (not far from the average total number of particles in a similar area of  the  experiments) interacting particles in a box of size $L_0$ with periodic boundary conditions. In Fig. 6 we show the simulation setup  with the symbols representing the fundamental lengths  of the simulations. In units of the sperm head's diameter $\sigma$ we have used $L_1=5$, $L_3=16$, $L_2=L_3/2+L_1=13$, and $L_0=64+\sqrt{2} L_{gap}$, while the width of the walls is $H=2$, and clearly the radius of the half-circle representing the curved part of the petals is $L_3/2$.


\begin{figure}
 \centering
 \includegraphics[width=8cm]{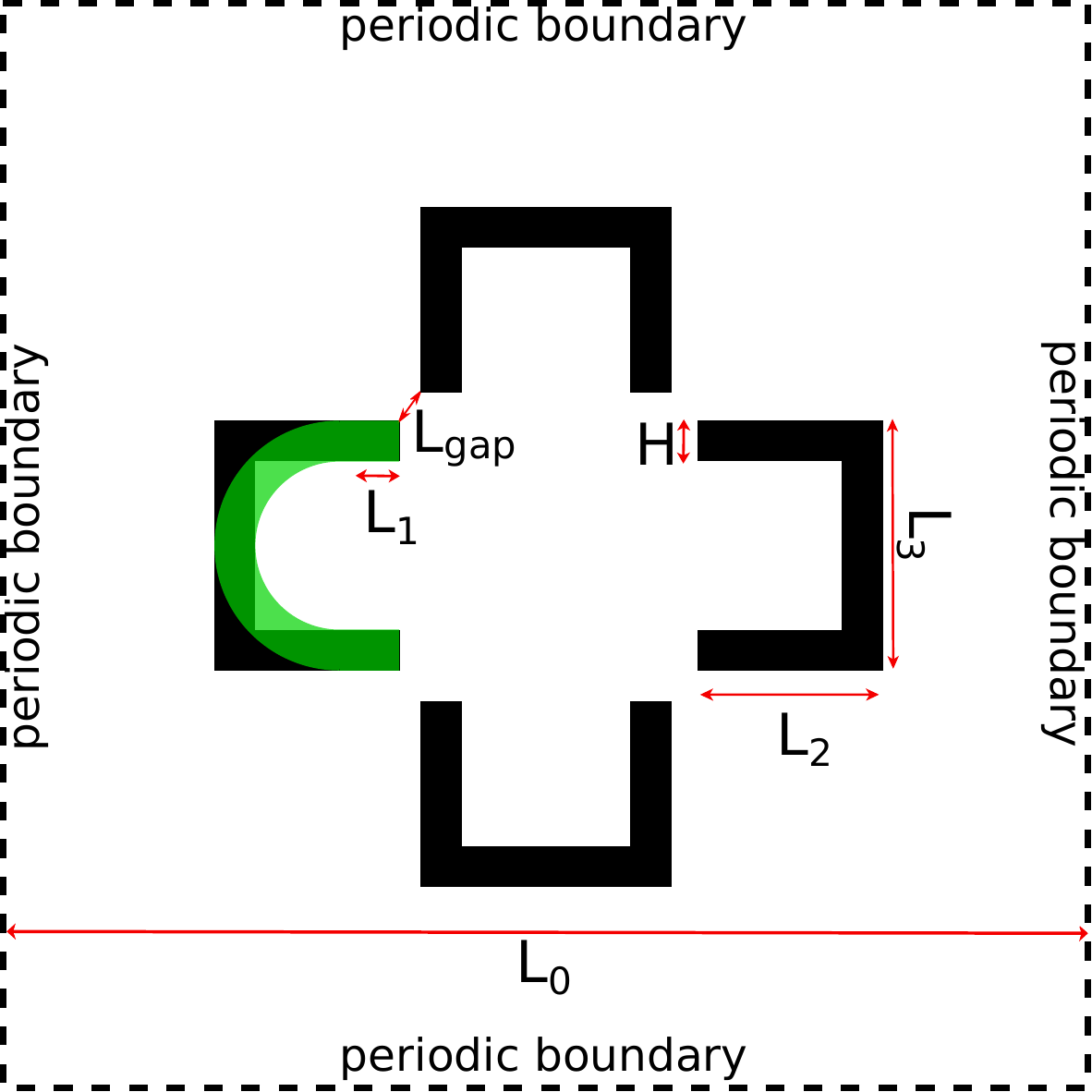} 
 \caption{{\bf Setup of the simulations whose results are presented in Fig. 4.} The sizes, given in units of $\sigma$ (the particle diameter, ideally corresponding to the head of the sperm cell) are $L_1=5$, $L_3=16$, $L_2=L_3/2+L_1=13$, and $L_0=64+\sqrt{2} L_{gap}$, while the width of the walls is $H=2$, and the radius of the half-circle representing the curved part of the petals is $L_3/2$. The value of $L_{gap}$  is varied between $1$ and $9$ (again in units of $\sigma$), as shown in Fig. 5.}
 \label{fgr:meth1}
\end{figure}

\section*{Author Contributions}

B. N., A.P., C. M., R. D. L. and I. V. conceived the idea and planned the experiments. I.V., A. Z. and V. A. designed and provided the microstructured devices.  C. M., B. N. and I. V. performed the experiments. C. M., B. N. and A. P. carried out the experimental data analysis. L.C. performed the numerical simulations and their analysis.  All the  Authors contributed  to the writing of  the manuscript. 

\section*{Conflicts of interest}
There are no conflicts to declare'.

\section*{Acknowledgements}
BN and AP warmly acknowledge Valentina Casciani, Matteo Verdiglione, Marco Toschi and Daniela Galliano (all from IVI Roma Labs) for useful discussions and suggestions about sperm thawing and observation protocols. BN and AP also acknowledge the financial support of Regione Lazio through
the Grant "Progetti Gruppi di Ricerca" N. 85-2017-15257 and from the
MIUR PRIN 2017 project 201798CZLJ.  LC acknowledges support from the
Alexander von Humboldt foundation. 
I.V. acknowledge dr. E. Quintiero for technical support and useful discussions.
AZ and VA acknowledge the Project "GENESI" - Development of innovative
radiopharmaceuticals and biomarkers for the diagnosis of tumors of the
male and female reproductive apparatus by the Italian Ministry of
Economic Development.



\bibliography{spermtrapping} 
\bibliographystyle{rsc} 

\end{document}